**Title**

**Pili mediated intercellular forces shape heterogeneous bacterial microcolonies prior to multicellular differentiation.**

Authors

Wolfram Pönisch[^,%], Kelly Eckenrode[#,$], Khaled Alzurqa[#], Hadi Nasrollahi[#], Christoph Weber[^], Vasily Zaburdaev[^,&,*] and Nicolas Biais[#,$,*].

Affiliations: [^]Max-Planck-Institute for the Physics of Complex Systems, Dresden, Germany, [#]Brooklyn College of CUNY, Department of Biology, Brooklyn, U.S.A, [$]Graduate Center of CUNY, New York, U.S.A.

Current addresses: [%] MRC Laboratory for Molecular Cell Biology, University City London, London, UK. [&]Friedrich-Alexander-Universität Erlangen-Nürnberg, Erlangen, Germany.

Corresponding Authors: Vasily Zaburdaev and Nicolas Biais

Brooklyn College and The Graduate Center

City University of New York

Biology Department, Room 307NE

2900 Bedford Avenue

Brooklyn, NY 11210

Phone: Office 718-951-5000 x6954

Email: nicolas@mechano-micro-biology.org

**Abstract:**

Microcolonies are aggregates of a few dozen to a few thousand cells exhibited by many bacteria. The formation of microcolonies is a crucial step towards the formation of more mature bacterial communities known as biofilms, but also marks a significant change in bacterial physiology. Within a microcolony, bacteria forgo a single cell lifestyle for a communal lifestyle hallmarked by high cell density and physical interactions between cells potentially altering their behaviour. It is thus crucial to understand how initially identical single cells start to behave differently while assembling in these tight communities. Here we show that cells in the microcolonies formed by the human pathogen *Neisseria gonorrhoeae* (Ng) present differential motility behaviors within an hour upon colony formation. Observation of merging microcolonies and tracking of single cells within microcolonies reveal a heterogeneous motility behavior: cells close to the surface of the microcolony exhibit a much higher motility compared to cells towards the center. Numerical simulations of a biophysical model for the microcolonies at the single cell level suggest that the emergence of differential behavior within a multicellular microcolony of otherwise identical cells is of mechanical origin. It could suggest a route toward further bacterial differentiation and ultimately mature biofilms.

**Introduction**

It is now broadly accepted that bacteria principally exist as surface-associated communities called biofilms [1,2]. Formation and survival of biofilms are a major concern, both in a medical and industrial context [3–5]. On the other side, biofilms can also provide useful applications for wastewater treatment [6] and are important for the proper functioning of many ecosystems [7]. The early stages of biofilm development are usually characterized by the formation of tethered small aggregates, so-called microcolonies, either by successive recruitments of new bacteria from the surrounding bulk fluid, multiplication of adhered bacteria or aggregation of bacteria actively moving on a surface [2]. Early microcolonies are comprised of dozens to thousands of cells, are often assembled in matter of hours and have been observed in many different bacteria species [8,9]. Microcolonies represent the first stage of a usually complex development into mature differentiated multicellular biofilms [1]. However, microcolonies are also commonly found by themselves *in vivo* [8,10–12]. For instance, a recent study has mapped out the advantage that microcolonies could represent for infection of the human pathogen *Neisseria meningitidis* [13]. Microcolonies thus also play an important role beyond that of an intermediate towards biofilm maturation. The first measurable step towards obtaining cells with different biological functions, the hallmark of differentiation, will be heterogeneous gene expression among the cell population. Understanding the nature of this transition from unicellular to multicellular lifestyle holds the promise for unraveling some of the mechanisms leading to differentiation and controlling biofilm development.

While eukaryotes and prokaryotes are obviously dissimilar, a parallel can be drawn between the development of bacterial biofilm and the development of complex multicellular organisms [2]. The cues that dictate the cascade of signaling that governs multicellular development have long been thought to be of chemical origin but recent studies have brought to light the importance of the mechanical interactions between eukaryotic cells in the differentiation process [14–16]. With the knowledge accumulated in the dynamics of eukaryotic cell aggregates in mind, we will explore here the internal dynamics of bacterial microcolonies.

In many bacterial species, the events leading to microcolonies are highly regulated with ultimately many different interactions governing the adhesion of bacterial cells to the substrate and between bacterial cells [17–19]. Often biofilms are held together by an extracellular matrix composed of DNA, excreted polysaccharides and various bacterial appendages [20,21]. In contrast, the formation of microcolonies of *Neisseria gonorrhoeae* (Ng) is solely relying on the interactions mediated by a ubiquitous appendage, the Type IV pilus (Tfp) [22,23]. Mutants lacking Tfp are not able to form microcolonies [24]. The unique reliance on Tfp makes Ng an ideal model system to fully understand the dynamics of formation of bacterial microcolonies. In this study, we look experimentally at the dynamics of formation of Ng microcolonies and highlight the crucial role of the mechanical forces generated by retractile Tfp in this process. Our central result is the discovery of emerging heterogeneous behavior within bacterial microcolonies within the first hours of formation. We observe a sharp gradient of bacterial motility from mobile surface layer towards nearly immobile bulk of the microcolony. These results are corroborated by experiments with bacteria incapable of Tfp retraction and comparison with the predictions of the *in silico* model we recently developed [25]. Ultimately, we see that heterogeneous gene expression follows the heterogeneous motile behavior.

**Results and Discussion**

**Ng microcolonies merge with dynamics consistent with a heterogeneous composition.** Tfp are retractile bacterial appendages whose cycles of elongation and retraction enable bacteria to exert forces on their surroundings [23,26]. These polymers have a diameter of molecular size (below 10 nm) and length exceeding the size of the bacteria body (several microns) [23]. An average Ng cell has 10-20 Tfps. Tfp can generate forces up to the nanonewton range when in bundles [27]. In the case of Ng, Tfp are the only motility appendage that the bacteria possess. This leaves the cycles of elongation and retraction of Tfp and the forces that Tfp can exert on their surroundings as the principal agents of microcolony formation. Ng bacteria can form nearly spherical microcolonies of upward to thousands of cells within a few hours, which greatly facilitates their study (See Figure 1a, Supplementary movie S1). The active merging of smaller microcolonies into a larger one is the central mechanism responsible for microcolony growth [24] (See Figure 1a,b,c, Supplementary movie S1). We took advantage of the fact that the merger of microcolonies necessitates a complex rearrangement of cells and thus will inform us on the internal dynamics of bacterial microcolonies. To this end, we studied in detail the dynamics of two merging microcolonies. Microcolonies were self-assembled by interacting bacteria. Microcolonies of the desired size could be subsequently retrieved and brought into close vicinity and let to interact under a microscope. To quantify the transition of two interacting colonies towards a spherical shape we used the images at the midplane cross section. By fitting an ellipse to the shape of the cross section we measured the aggregate's short and long symmetry axis. Their ratio approaches 1 from below as the colony rounds up (See Figure 2a and Figure S1 a,c). Additionally, we measured the height of the "bridge" – a contact area forming between the two touching microcolonies (See Figure 2a and Figure S1 a,b). Our analysis shows that the merger occurs with three different dynamical regimes. In the first rapid regime (a few seconds) the two microcolonies are pulled together by retracting pili (retracting speed of Tfp is of the order of 1 $\mu$m/s and hence the time scale). In the intermediate regime (a few minutes) the two microcolonies smoothen the gap in the contact area and attain an ellipsoidal shape. Finally, the slowest regime (half an hour to more than one hour) is where the two microcolonies round up to a sphere (See Figure 2b and Figure S2). To follow the mixing of cells during merger we used fluorescently labeled bacteria creating microcolonies with two different colors (expressing either YFP or tdTomato fluorescent proteins). The merger showed a mostly flat contact region (See Figure 2c and Supplementary movie S2). These findings would be consistent with a heterogeneous microcolony where an envelope of mostly motile cells drives the relaxation in the area with highest surface curvature (at the contact line between microcolonies) and is resisted by a core of much less motile cells responsible for the longer relaxation time. To probe directly for the existence of such "liquid-like" envelope and a more "solid-like" core, we decided to follow the motility of single cells as a function of their position in a microcolony.

**Ng microcolonies have an outer layer of motile cells and a core of far less motile cells thus presenting a heterogeneous motility behavior.** To look at the motility of single cells within microcolonies we used a mixture of wild-type (WT) non-labeled cells mixed with a small percentage (5 to 10%) of fluorescently labeled bacteria. We used fluorescent and bright field channels to simultaneously record bacterial tracks and determine their location within the microcolony. To quantify the motility of cells in the microcolonies without being hindered by the global motion of the microcolony, we evaluated the changes in relative distance for pairs of cells at a given depth from the colony surface. The mean squared relative distance increments (MSRD) as a function of time provide information about the motility of cells (See Figure 3 a,b, Supplementary movie S3 and Methods section) [28]. For short time scales (<1minute) and for all depths the MSRD grows linearly with time, consistent with normal diffusive behavior of cells. However, the intensity of the random motion (quantified by the value of the diffusion constant) declines sharply for bacteria at different depth away from the microcolony surface (deeper into the microcolony) (Figure 3 c,d). These results indicate that Ng bacteria as they assemble into microcolonies create a microenvironment with strikingly different motility behaviors. In order to investigate the nature of the emergence of differential motility within a microcolony we turned to computer modeling. We use a numerical simulation of a bacterial microcolony at the single cell level of detail with explicit pili-pili interactions [25]. This model accounts for known Tfp dynamics and force generation, along with the physical shape of cells and reaches the limits of current computational time constraints (See Methods section and Supplementary Information). The simulations recapitulate the motility behavior observed experimentally: the mean square

displacements of bacteria pairs show diffusive motility throughout a microcolony at short time scales, but the value of the associated diffusion constant diminishes greatly for the bacteria tracked deeper towards the core of the colony (See Figure 3 e,f). Importantly, since all cells in our computational study have identical features, the emergence of the heterogeneous behavior highlights the importance to take a dynamical approach when analyzing any aggregation of cells. An energetic approach that would only take into account the difference in adhesion properties between the cells will most likely miss the complex reality that a more precise and dynamical approach might uncover [29]. In the simulations we can correlate the heterogeneous motility of cells with an average number of Tfp bound to pili of other cells and the times of presence and attachment of each pilus. The number of attached pili slowly decreases going from the core of the microcolony to the outer layer as the fluctuations in the number of pili (attached or not) increases (See Figure S3 a and b). Additionally, in the simulations the attachment times rapidly decrease when going from the core to the outer layer (See Figure S3 c). For the cells deep in the core of the colony, even if one of their pili detaches, the multitude of remaining contacts holds the cell in place long enough for the detached pilus to bind again, thus leading to an effectively very slow dynamics in the core. Due to a higher cell density compared to the colony surface, within the bulk the pili density is increased, thus increasing also the re-binding rate of a pilus. In other words, cells in the core are more likely to be bound to each other than cells on the outside. The qualitative agreement between experiments and simulations indicates that the occurrence of different motility behaviors within a microcolony can be explained solely by the intercellular forces exerted by Tfp. If we use the same model to determine the dynamic of microcolonies merger we obtain similarly good qualitative agreement with the experiments (See Figure S4). Our model contains just a few free parameters with the other parameters provided by literature (Supplementary data Table I), however, while reproducing the data qualitatively it still deviates in numerical values. One observation is that our *in silico* colonies are less heterogeneous as compared to experiments with a sharper transition. Most likely the discrepancy is due to the simplification of the model allowing for one pilus to have at maximum one contact point with any other pili of other cells. While dictated by computer feasibility it leads to more "dynamic" microcolonies as compared to experiments. Multiple pili contact points in the real setting lead to the formation of active pili network with higher gradient in motility and less dynamic core of the cells. It is the interesting direction of further research to understand the biophysical properties of such a network. Recently a different modeling approach averaging the contribution of all pili to an intermittent attractive force[13] similarly led to the existence of a smooth diffusion gradient stressing emphasizing the importance to be able to take into account multiple pili interaction in the future. Importantly, we see that the heterogeneous behavior in the microcolonies emerges during the assembly process of physically identical bacterial cells and this phenomenon is very robust for a wide range of model parameters (see [25] and Supplementary Information).

**Non-retracting cells are excluded on the outside of microcolonies.** If the forces due to retracting Tfp play the central role in shaping the microcolony, it would be predicted that cells deprived of the ability to generate those forces would alter the process of colony formation. To test this in the model, we performed simulations of a 1:1 mixture of pulling wild-type cell and cells that possess pili but are unable to retract them. We observed that the non-pulling cells were excluded to the outside of the microcolonies (See Figure 4c). To find the actual experimental confirmation of this prediction, we used the fact that retraction of Tfp in Ng is under the control of a AAA ATPase pilT [30]. Bacteria deprived of this molecular motor (NgΔpilT) still have Tfp and thus can interact through Tfp-Tfp interactions with other bacteria but they do not retract their pili and thus cannot exert forces on their surroundings [22]. A 1:1 mixture of NgWT cells and NgΔpilT (fluorescently labeled) led, in agreement with the prediction of the computational model, to the sorting of cells that cannot exert forces to the outside of the microcolonies during the self-assembly process (See Figure 4 a,b). Similar sorting of two types of bacteria were observed previously in the case of bacteria with different types of pili and thus different interaction forces between them [31]. Here, the pili are made of the same major pilin in both, pulling and non-pulling bacteria. Thus, the origin of the sorting in our experiments is the absence of retraction forces, not the difference in the interaction forces between different types of pili. A similar discrepancy exists in eukaryotic systems where the difference of adhesion forces between cells has been first postulated to be a driving force in early embryogenesis [32]. More recent theories take also into account the ability of eukaryotic cells to exert forces on their surrounding through contraction of their skeleton. In this case the dynamics of the cells seem crucial to the biological outcome [29,33]. In a control experiment, when the fluorescent markers were used to label two NgWT populations with different fluorophores in a 1:1 mixture, the corresponding forming microcolonies showed a homogeneous repartition of both types of bacteria (See Figure 4 d,e). The homogeneous repartition is also obtained in the case of the simulations (See Figure 4 f). These results exemplify the crucial role that Tfp contractile forces have in shaping Ng microcolonies and demonstrate the predictive power of the computational model.

**Heterogeneous gene expression follows heterogeneous motility behavior in microcolonies.** We have demonstrated that Tfp retraction forces in a set of initially identical cells can lead to the formation of a heterogeneous motility behavior as they form a microcolony. To probe whether this motility behavior is associated with differential gene expression across a microcolony we have generated a gene reporter for the pilin gene: the promoter of the *pilE* gene was incorporated heterologously in the genome of Ng driving a mCherry fluorescent protein (Ng *pilE::mCherry*) to get an idea of the expression of pilE across a microcolony. The genomic site of incorporation has been shown to not modify the behavior of the bacteria and the fluorescence will be a proxy for the expression of the pilE gene. When these microcolonies were treated in similar conditions as the ones we used to study the dynamics of microcolonies, the fluorescence reaches the edge of the microcolony and the fluorescence is consistent with a homogeneous expression across the microcolony (See Figure 5a). Our measurements of metabolic levels among cells within microcolonies during the first few hours of formation show also homogeneous profiles (See Figure S5 a,b). As we have seen previously in the case of Tfp numbers, life and attachment times, the computer model enables us to access quantities that are difficult to measure experimentally. Importantly, we can also compute the mechanical forces within the microcolony. According to the model the dynamical formation of bacterial microcolonies via Tfp interactions will lead not only to a gradient of numbers of Tfp and times of attachments (See Figure S3 a, b and c) but also to the appearance of a force gradient across a microcolony (See Figure 5c). So the heterogeneous motility within microcolonies is linked to a force gradient. After a few hours of interactions, the force gradient is present, but the gene expression and metabolism are homogeneous (See Figure S5 a,b and Figure 5a). All small diffusible chemical cues are most likely also spatially homogeneous due to the small size of the microcolonies. For bacterial aggregates below 40 microns in diameter diffusion is a very efficient process and both simulations and experiments indicates that oxygen, nutrients and waste products will be homogeneously distributed [34,35].

When microcolonies are left to develop for 7 hours in liquid the pattern of gene expression of pilE becomes heterogeneous as shown by the fluorescence across the microcolony (See Figure 5b). The pattern of gene expression parallels the pattern of differential motility that we observed. But importantly pattern of gene expression appears multiple hours after the observation of the motility pattern and the metabolic activity is still homogeneous at that later 7 hour time point (See Figure S5 a,b). These results are indicative that an initial change of motility behavior precedes a differential gene expression within the microcolony. The pursuit of the mechanisms relating these gradients to differential gene expression in a microcolony is beyond the scope of the present study. Nevertheless, keeping in mind the idea from eukaryotic mechanobiology that a gradient of mechanical forces can trigger difference in gene expression, it is tantalizing that the gradient of forces within a bacterial microcolony triggered by the dynamics of Tfp can be the first step towards differentiation as it precedes heterogeneity in gene expression.

**Conclusion**

The development of multicellular eukaryotic organisms has been a long standing biological question with an overwhelming focus on the presence or gradients of different molecules until recently, when the role of physical forces in development was acknowledged [14]. Bacterial biofilms have been recognized as being similar to multicellular entities and being able to develop differentiated state usually over multiple days [1,2]. The mechanisms at play in this development are the subject of intense scientific inquiry due to the health related importance and ubiquity of biofilms. The role of mechanics in bacteria lifestyle, in particular within biofilm, is being recently appreciated, whether it is the role of motility, hydrodynamical flow or internal forces building within biofilms [36]. Besides, there is now mounting evidence that bacteria have the ability to sense mechanical forces and those forces can in turn change genetic expression [37,38].

In the case of *Neisseria gonorrhoeae*, we have shown that a group of identical cells powered by cycles of extensions and retractions of Tfp can self-organize in microcolonies with heterogeneous motility pattern. Such heterogeneous motility behavior correlates with a gradient of mechanical forces within a microcolony. If these mechanical forces can be sensed by bacteria, a feedback mechanism could provide the ability to these forces to take over the control of the spatial development of the microcolony by triggering heterogeneous gene expressions. Ng microcolonies are akin to early developing embryos and eukaryotic cell spheroids, where the forces between cells, both adhesive and contractile, play a crucial role [15,33,39–41]. What might unify these biologically so different systems is the common biophysical mechanism where the mechanical forces generated on the surface of aggregates drive the shape transformation which is resisted by the viscoelastic response of the bulk. This mechanism can apply generally despite the difference in origin of these aggregates whether it is motile aggregation of different cells or successive divisions of a starting cell or a combination of those two modalities. The simple system of Ng bacteria, accompanied by the *in silico* model, not only enables us to understand better the physiology of an important human disease but it could also give a new insight into the earliest steps of genetic differentiation within a group of identical bacterial cells and ultimately the evolution of multicellularity [42].

**Methods**

**Bacteria strains and growth conditions**

The wild-type (WT) strain used in this study is MS11. The ΔpilT mutant was obtained by an in-frame allelic replacement of the pilT gene by a Kanamycin resistance cassette. Fluorescent proteins (YFP, mCherry or tdTomato) driven by a consensus promoter were incorporated by allelic replacement together with an antibiotic marker (either Kanamycin or Chloramphenicol). Similarly, mCherry driven by the reporter of the pilin gene (370 bp before the beginning of the starting ATG of the pilin ORF) was incorporated by allelic replacement together with a Chloramphenicol marker. Bacteria were grown on GCB-medium base agar plates supplemented with Kellog's supplements at 37°C and 5 % $CO_2$. 80 μg/ml of Kanamycin or 7 μg/ml of Chloramphenicol were added when growing mutants with the corresponding antibiotic resistance cassette. Cells were streaked from frozen stock allowed to grow for 24 hours and then lawned onto identical agar plates and used after a 16 to 20 hour growth period. The list of the strains used in this study can be found in the table below:

| Strain Name | Fluorophore | Antibiotic marker | Reference |
|---|---|---|---|
| **WT MS11** | None | None | [43,44] Gift from M.So |
| **WT YFP** | YFP under a consensus promoter | Chloramphenicol | *This study* |
| **WT mCherry** | mCherry under a consensus promoter | Chloramphenicol | *This study* |
| **WT tdTomato** | tdTomato under a consensus promoter | Chloramphenicol | *This study* |
| **ΔpilT mCherry** | mCherry under a consensus promoter | Chloramphenicol and Kanamycin | *This study* |
| **WT *pilE::*mCherry** | mCherry under the pilE promoter | Kanamycin | *This study* |

**Microcolony formation**

Bacteria from lawns on agar plates were resuspended in 1 ml of GCB medium at an optical density of O.D.=0.7. 100 μl of the suspension was added in the well of the 6 well plate containing 2 ml of GCB medium with a BSA coated coverglass (round 25 mm diameter coverglasses (CS-25R) Warner Instruments) at the bottom or without. The 6 well plate was centrifuged at 1600xg in a swinging bucket in an 5810R centrifuge (Eppendorf) for 5 minutes resulting in single bacteria uniformly coating the bottom of the well. For direct imaging the coverglass were transferred to an observation chamber (Attofluor cell chamber, Thermo Fisher Scientific). In the case of mixture the suspension at O.D.=0.7 of both components of the mixture were prepared and a new 1 ml was prepared by the proper ratio of the two suspension. 100 μl of that new suspension was used similarly to what was described previously.

**Microscopy and microcolony merger.**

All movies were obtained on a Nikon Ti Eclipse inverted microscope equipped for epifluorescence and DIC microscopy and with an optical tweezers setup all under an environmental chamber maintaining temperature, humidity and $CO_2$ concentration. The objective used is a 60X plan Apo objective. The camera used were either a sCMOS camera (Neo, Andor) or a CMOS USB camera (DCC1240M, Thorlabs). 1Hz fluorescent movies and 0.1Hz DIC movies of either microcolony merger or follow up of single cell motility were taken for further analysis. In the case of microcolony merger, microcolonies were preformed and were brought into contact either by optical tweezers or hydrodynamical flow.

**Scanning Electron Microscopy**

Bacteria microcolonies on a glass coverglass were fixed with 3.7% formaldehyde in PBS pH 7.4 for one hour. The microcolonies were subsequently washed 3 times with PBS and then dehydrated by step in ethanol (50%, 70%, 80%, 90% and 100% ethanol). The samples were then critical point dried and imaged on a Hitachi S-4700 Scanning Electron microscope.

**Image analysis of experimental data**

*General information.* We analyzed in total 28 merger events and 40 individual colonies for the tracking of individual cells (with at least 42 trajectories used for the computation of the spatially dependent MSRD) inside of microcolonies. Matlab R2015b was used for edge detection of DIC movies and tracking of individual cells inside of colonies.

*Edge detection.* In order to detect the edges of single colonies and the merger from DIC data the same algorithm was used. We computed the first derivatives of intensities in x- and y-direction of a Gaussian filtered image and thresholded its absolute value. Afterwards we dilated and eroded the binary image, filled all remaining holes and removed small objects. For all steps internal functions of Matlab were used.

*Single Cell Tracking inside of Colonies.* To track single cells from the fluorescence images, we first computed the center of mass (COM) of the binary shape resulting from the edges and corrected the fluorescence data in order to reduce the effects of the translations of the colony on the tracking algorithm. We interpolated the position of the COM from the DIC data recorded with frequency of 0.1 Hz to match that of the fluorescence data with 1 Hz resolution. A cubic spline data interpolation was applied on the x- and y-component on the COM. We next used the detection and tracking algorithm developed by Blair et al[45]. For original images we computed the background by applying a large-scale Gaussian filter and substracted its values from the images. Additionally, we used a smaller Gaussian filter for smoothing.

*Estimating the MSRD and the diffusion coefficient of cells inside of colonies.* The measurements of displacement of individual cells is strongly affected by translation and rotation of the microcolony as a whole. This effect can be circumvented by measuring the absolute value of the distance vector of two individual cells and computing the time-averaged second moment of the relative distance increments, as a function of the time lag. One can show that this quantity is equal to the sum of the mean squared displacements of the two individual cells in the colony [Note: The ensemble averaged mean squared relative displacement coincides with the time averaged quantity as long as the individual cells displacements are smaller than the initial distance between the cells, which is the case for most of pairs of cells in our measurements.]. It thus grows linearly with the lag time and the prefactor of this linear growth quantifies the diffusion of the cells. To identify the location of cells in the microcolony, we computed the distance from the surface by finding the edge shape from DIC images taken in parallel to the fluorescent images and picking the cell pairs that move in a region. By assuming that cells belonging to the same region have the same diffusion coefficient we can read out of individual cells from the fit. The constant offset $b$ accounts for the measurement noise. This method allows us to find the diffusivity of cells as a function of the distance from the surface.

*Ellipse fitting.* The binary images of the merger movies allowed us to fit an ellipse by computing the central moments, as explained in[46]. This algorithm also allowed us to compute the orientation of a non-spherical colony, the length of the short and long axis of the ellipse and their axis ratio (Figure 2b).

*Bridge height measurement.* To compute the height of the bridge forming between the two regions of the binary images we rotated the image so that the colonies were oriented along the x-axis and calculated the COM of the combined regions. Then we moved a line of length $L$ centered around the COM and perpendicular to the axis connecting the two colonies. The bridge was defined as the range for which the whole line could be found inside of the colony region. *L* was chosen to be small enough to be not affected by the elliptical shape of the colonies. For the late coalescence the results were compared to the short axis of the ellipse.

**Simulations**

*General information.* We implemented the simulation of the model (see Supplementary Information for a summary of the model) in C++ and used the package OpenMP for parallelization on up to eight CPUs. The simulations were performed on the local computing cluster of the Max Planck Institute for the Physics of Complex Systems consisting of x86-64 GNU/Linux systems.

*Single colony modeling.* For the simulation of the dynamics of individual cells inside of a colony we randomly initialized 1700 cells inside of a sphere without cell-substrate interactions. First we allowed the cells to repel each other until there was no overlap between neighboring cells. Next, we introduced pili and their dynamics, causing the motion of cells. In order to reproduce the experimental results, we only monitored the cells being positioned <1 μm above and below the midplane and computed the second moment of the relative distance increments (as for the experimental data) of cell pairs, projected to the midplane.

*Merger modeling.* For the merging simulations we again neglected cell-substrate interactions and initialized two separate colonies each consisting of 1000 randomly distributed cells. The spherical shape of microcolonies indicates cell-substrate interactions are negligible as, if not, they would deform the microcolony towards the substrate [25].For the first 100 s the pili of the colonies were not allowed to interact with pili of the other colony. This way we allowed for the formation of stable individual microcolonies with a radius of ~ 7 μm and initial separation of 16 μm between their centers. To compute the bridge height and the eccentricity we created the binary image from the projection of the cell positions and their shapes onto a plane tangential to the axis between the two colonies.

*Assembly modeling.* For the assembly experiments we distributed 1500 cells on a substrate of the size 98.56 x 98.56 $\mu m^2$ with periodic boundary conditions. The pili of a fraction of cells were not able to retract, modeling the ΔpilT mutant.

**Acknowledgements**

We would like to acknowledge the technical help of Luis Santos and Ingrid Spielman for the experiments, Frank Jülicher, Stefan Diez, Hugues Chaté, Guillaume Dumenil and Yen Ting Lin for fruitful discussions on the model and Guido Juckeland for his help with setting up the numerical model. N.B would also like to acknowledge funding from the NIH grant AI AI116566. W.P. kindly acknowledges support from the IMPRS-Celldevosys (International Max Planck Research School for Cell, Developmental and Systems Biology).

**Author contributions**

W.P., V.Z and N.B designed the experiments and project. W.P., K.A, K.E and H.N. performed experiments. W.P., C.W. and V.Z. designed the model and implemented the simulations. W.P., C.W. and K.E. analyzed the data. W.P and C.W. performed the simulations. W.P., V.Z and N.B wrote the paper.

**Competing interests**

The authors declare no competing interests.

**Figure Legends**

**Figure 1** *Tfp mediated microcolony mergers are a common mode of microcolony formation.* **a)** DIC micrographs of a timeseries showing two microcolonies merging among many interacting bacteria. See also Supplementary movie S1. Scale bar = 8 µm. **b)** Scanning Electron Micrograph of two merging microcolonies. Scale bar = 8 µm. **c)** Scanning Electron Micrograph of the bridge formed between two merging microcolonies. Scale bar = 4 µm.

**Figure 2** *Merging of Ng Microcolonies.* **a)** Merger of *Ng* microcolonies recorded with a DIC microscope (Scale bar = 10 µm). The red line highlights the detected edges. **b)** In order to estimate the time scales of the merging of two colonies the time-dependent bridge height and the symmetry axis ratio of a fitted ellipse were measured from the binary images. By fitting a function of the form $h(t) = h_0 \cdot \left(1 - \alpha \cdot \exp{-t/\tau_1} - (1-\alpha) \cdot \exp{-t/\tau_2}\right)$ to the bridge height we were able to estimate the first time scale corresponding to the initial approach of the two colonies and the time scale characterizing the closure of the bridge. The third time scale resulted from a fit to the aspect ratio of the short axis and the long axis of the ellipse and corresponds to the relaxation of the ellipsoidal colony to a spherical shape. **c)** Merger of two fluorescently labeled microcolonies. See also Supplementary Movie S2.

**Figure 3** *Motility of single cells inside a microcolony.* **a)** Representation of the detection of fluorescently labeled cells. The left image highlights the detection the fluorescently labeled cells. The right image shows the position of individual cells relative to microcolony. Scale bar = 10 µm **b)** In order to be able to reduce the effect of rotations of the microcolonies on the trajectories of single cells, we computed the mean squared relative distance of cell pairs. Both cells were defined to be a pair if they could be found in a similar region, defined by their distance from the surface. **c,d)** Diffusion coefficient $D$ from the experimental data as a function of the distance $R_s$ from the surface and MSRD as a function of time. **e,f)** Diffusion coefficient $D$ from the simulation data as a function of the distance $R_s$ from the surface and MSRD as a function of time.

**Figure 4** *Demixing of Ng Microcolonies.* **a)** DIC and fluorescence images allowed to detect the positions of the WT cells (right) and the ΔpilT mutants (center)**. b)** Intensity profile of fluorescently labeled WT (YFP) and ΔpilT mutant (mCherry) cells along a line in the midplane going through the center of the colony (a). **c)** Simulated assembly of a mixture of WT and ΔpilT mutant cells. The inset shows the midplane of a colony. The green cells are WT cells, the red cells are ΔpilT mutants. **d-f)** Same data as a-c for a mixture of differently labeled WT cells.

**Figure 5** *Heterogeneous genetic expression within a microcolony* **a)** Brightfield (left), fluorescence (center) and overlayed image (right) of a Ng *pilE::mCherry* microcolony after 3 hours of formation. (Scale bar = 10 µm.) **b)** Brightfield (left), fluorescence (center) and overlayed image (right) of a Ng *pilE::mCherry* microcolony after 7 hours of formation. (Scale bar = 10 µm.) Note that a) and b) represent two microcolonies imaged in the same conditions showing the spatial heterogeneity of expression in b) and not in a). **c)** Forces as a function of the distance of the center (COM) of an *in-silico* colony. $F_{ij}$ are the excluded volume force acting on a cell due to neighboring cells (blue) and the absolute values of pili forces acting on one cell (red) and where $i$ and $j$ are pili indices.

**Figure 1**

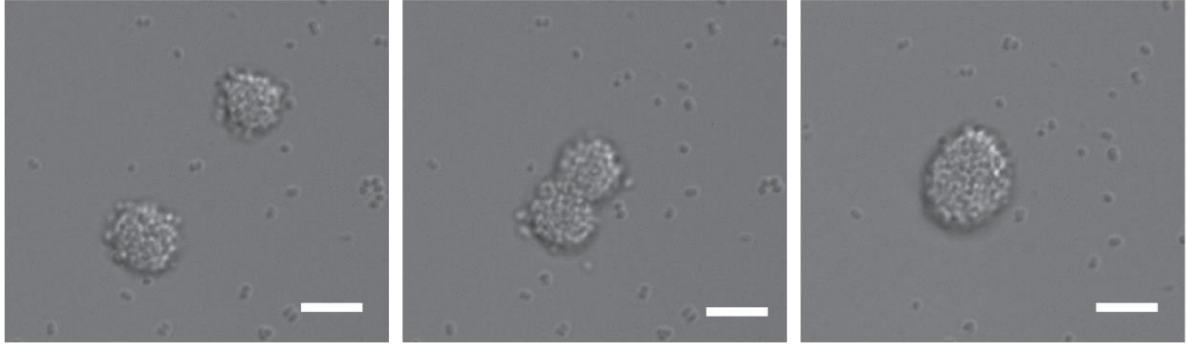

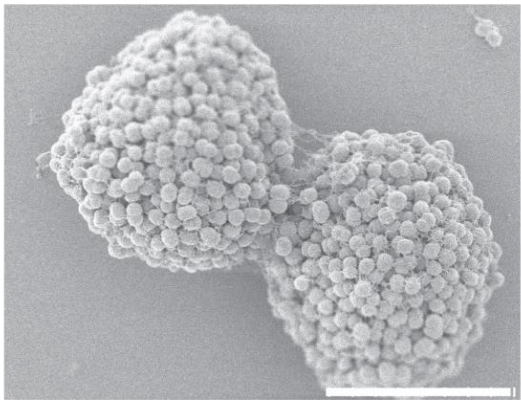

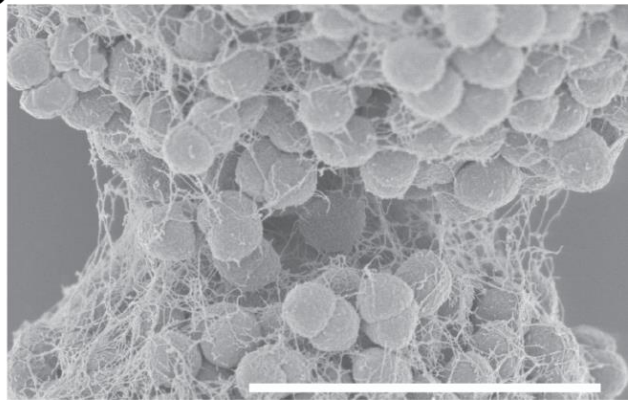

**Figure 2**

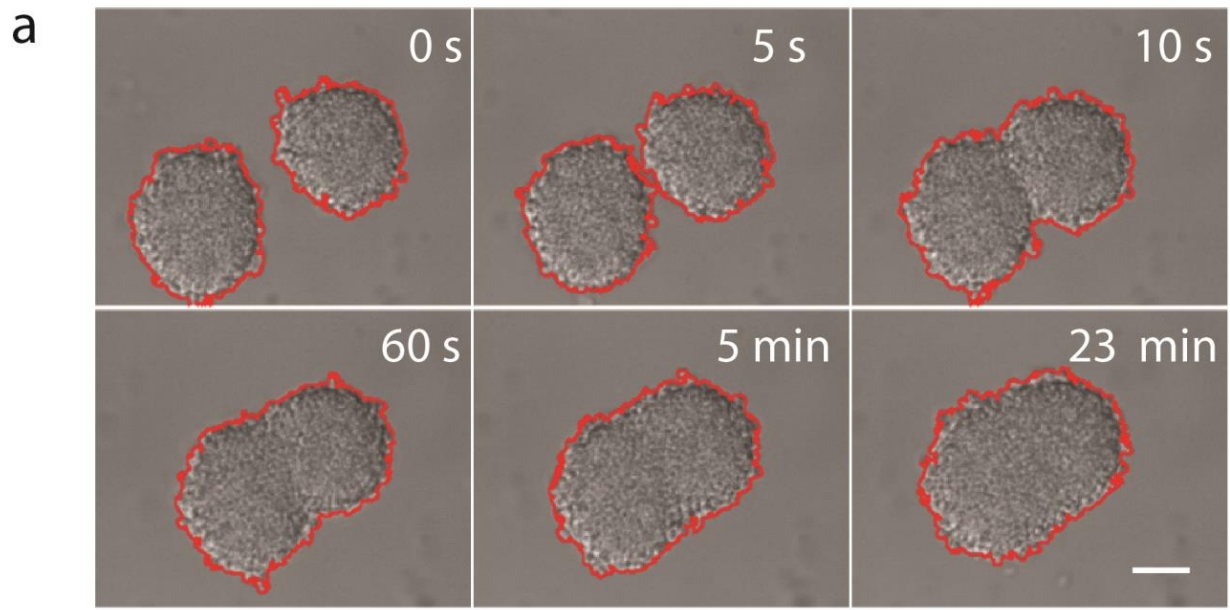

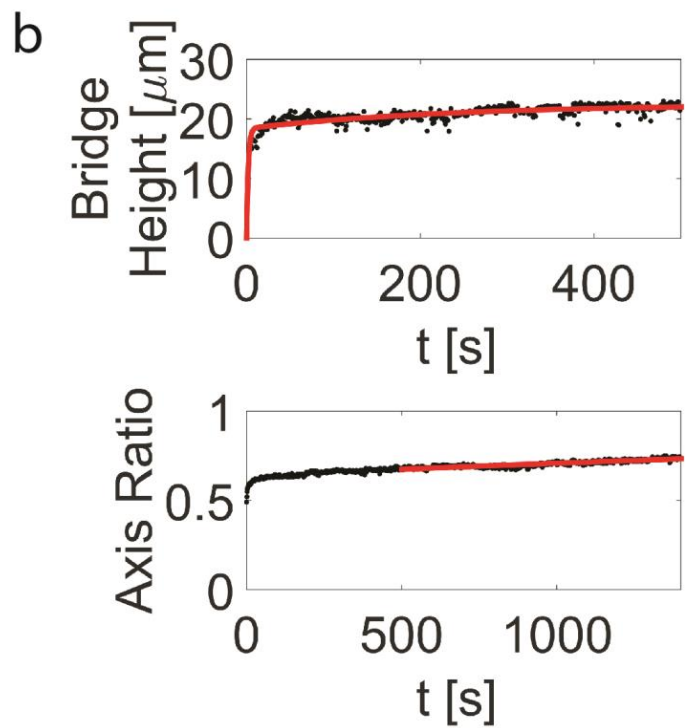

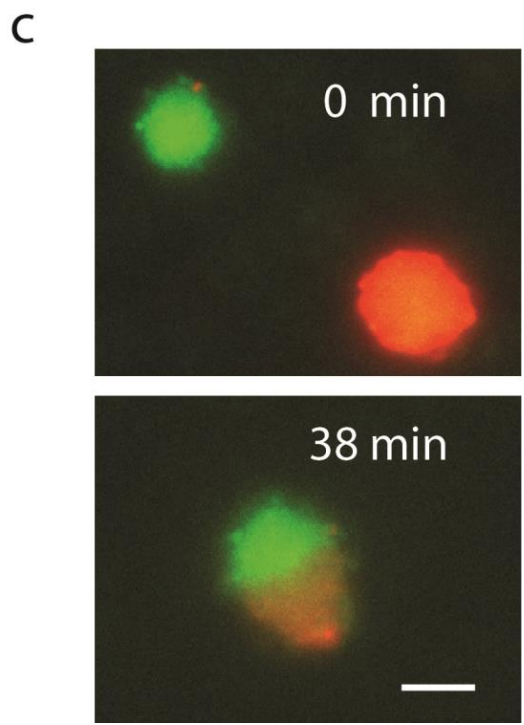

**Figure 3**

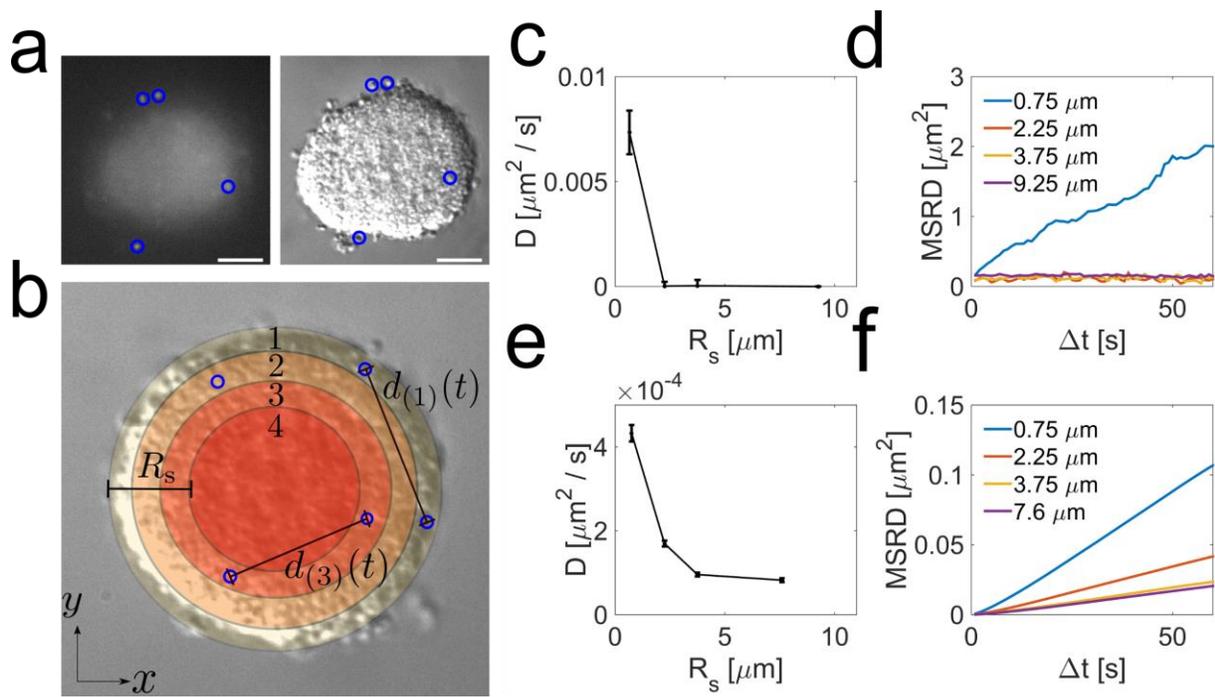

**Figure 4**

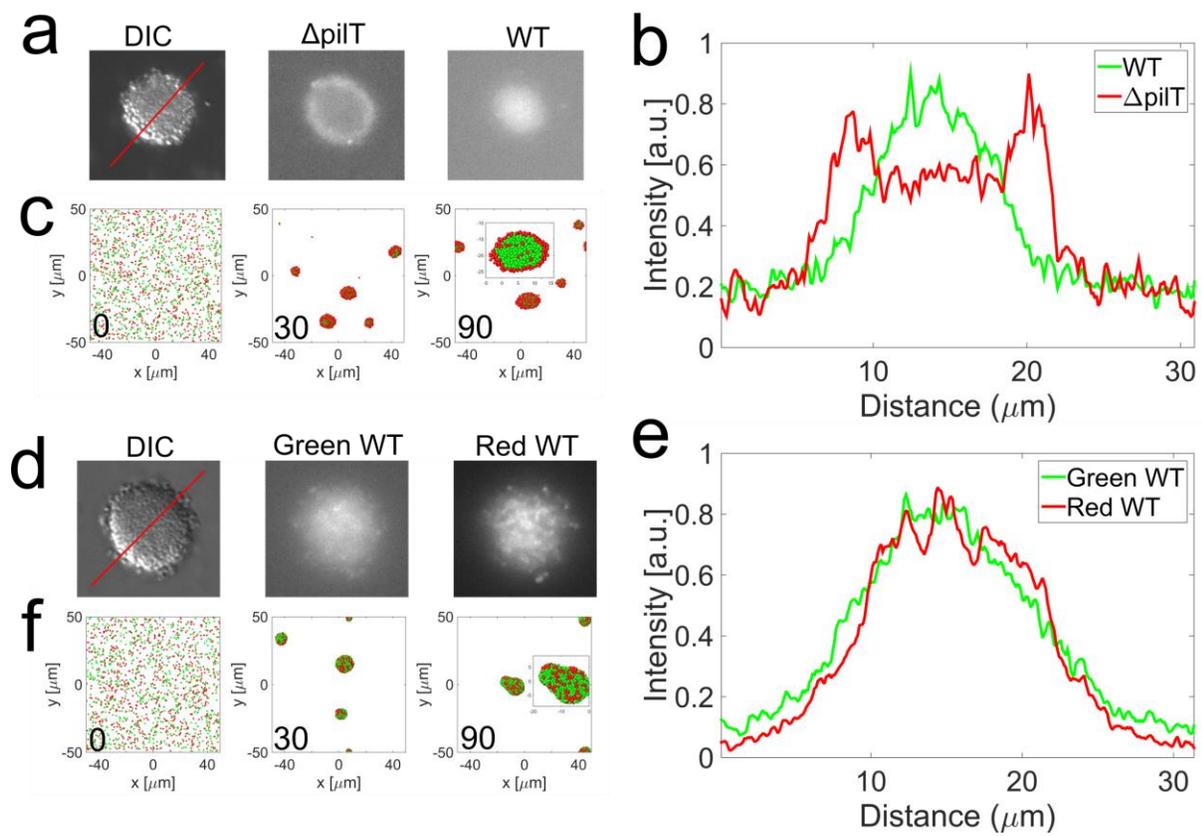

**Figure 5**

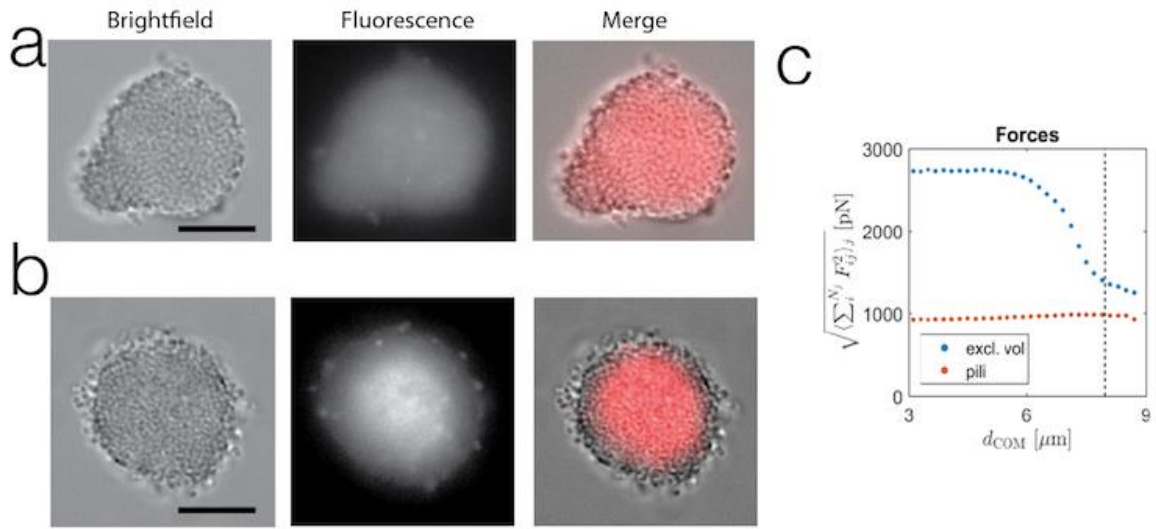

# Supplementary Information for

# Pili mediated intercellular forces shape heterogeneous bacterial microcolonies prior to multicellular differentiation.

Wolfram Pönisch, Kelly Eckenrode, Khaled Alzurqa, Hadi Nasrollahi, Christoph Weber,

Vasily Zaburdaev and Nicolas Biais.

**Methods:**

**Metabolic activity in WT Ng microcolony.**
In order to assess the metabolic activity of bacteria within a microcolony we used two different dye which fluorescence correlates with different aspect of cellular metabolism[1,2]. We used fluorescein diacetate (FDA) (known to measure esterase activity) and 5-cyano-2,3-ditolyl tetrazolium chloride (CTC) (known to measure dehydrogenase activity). In both cases bacteria were allowed to form microcolonies similarly to what as been described in the main methods. Then they were treated differently:
*FDA:* Stock solution of FDA were made by resuspending 5mg of FDA in 1 ml of acetone and stored at -20°C. After microcolony formation for 3 hours 10 µl of FDA was resuspended in 500ul of PBS. After careful removal of the GCB medium, the FDA PBS mixture was added to the chamber and incubated in the dark for 10 minutes. The chamber was then washed 5 times with PBS and the microcolonies were then imaged by fluorescent microscopy using a GFP filter set.
*CTC:* Stock solution of CTC were made by dissolving the dye in pure water at 33mM and stored at -20°C. After microcolony formation for 3 hours, the GCB medium was carefully removed and 500ul of 3.3mM CTC in PBS was added to the chamber and incubated at room temperature for 30 min. The chamber was then imaged by fluorescent microscopy using a Texas Red filter set.

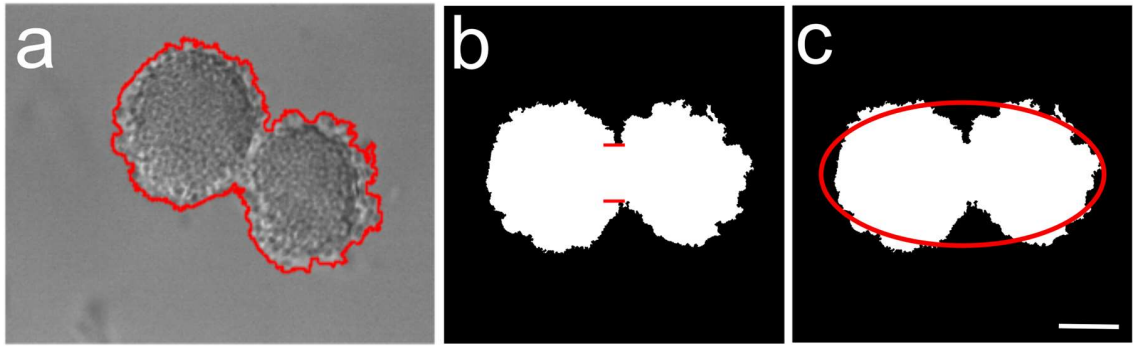

**Figure S1** *Analysis of Merger dynamics from experiments* – **a)** Result of the Edge Detection of the merger of microcolonies. **b)** After rotation and edge detection the bridge was measured as explained in the methods part. **c)** Ellipse fitted to the binary image of two colonies. (Scale bar = 10 µm).

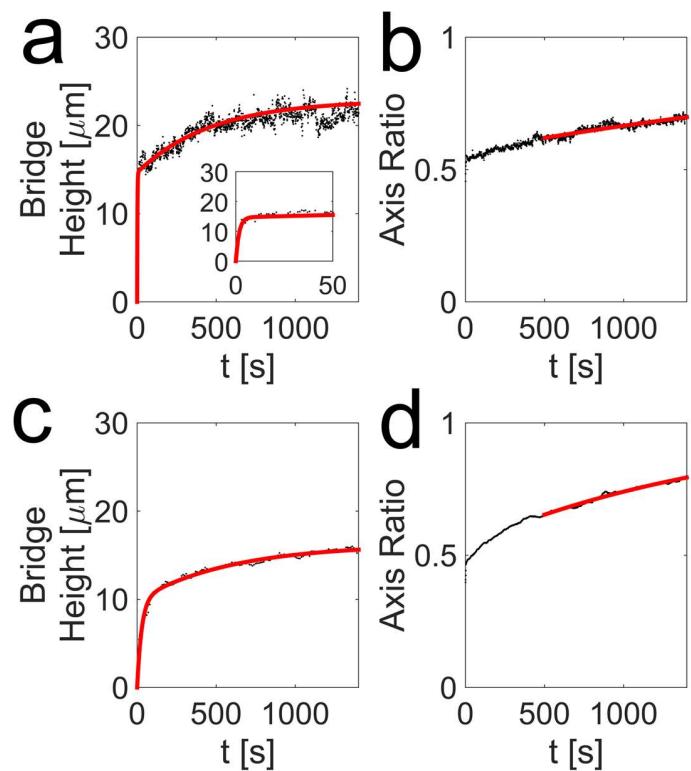

**Figure S2** *Time Scales of Merger* – **a)** Bridge height for experimental data. By fitting a function of the form presented in Figure2 we were able to measure the two first time scales. **b)** Axis Ratio for experimental data. By fitting we were able to compute the time scale of relaxation to a spherical shape. **c)** Bridge height for simulated merger. **d)** Axis ratio for simulated merger.

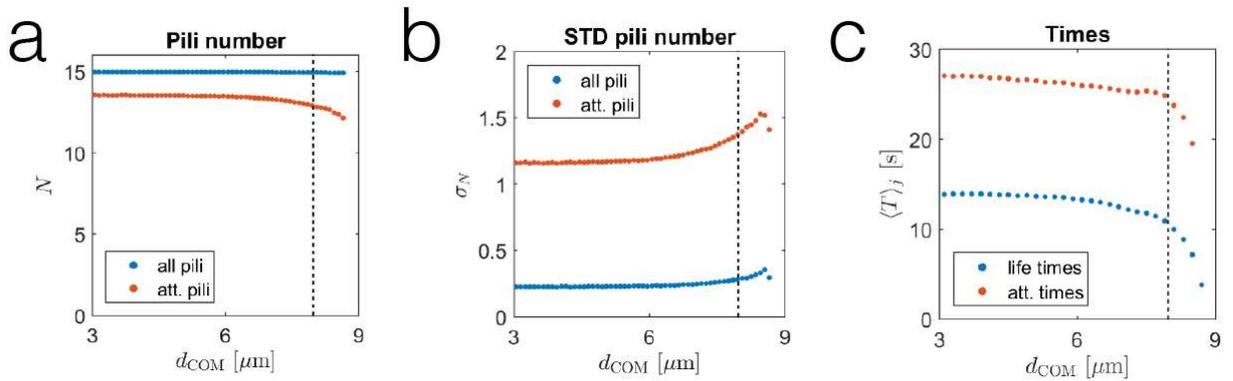

**Figure S3** *Number of Pili, life and attachment times for pili across a microcolony* **a)** Mean number of attached pili as a function of the distance from the center of a microcolony consisting of 850 cells. Close to the surface of the colony less pili are attached to other pili. **b)** Standard deviation of the number of attached pili as a function of the distance from the center of a microcolony consisting of 850 cells. Near the surface of the colony the fluctuations are stronger. **c)** Life times and attachment times of pili a function of the distance from the center of a microcolony consisting of 850 cells. Note that the mean life time of pili has a smaller value than the average attachment time. This discrepancy is explained by the exponential length distribution of pili, where shorter pili are less likely to bind at all and thus have a shorter life time.

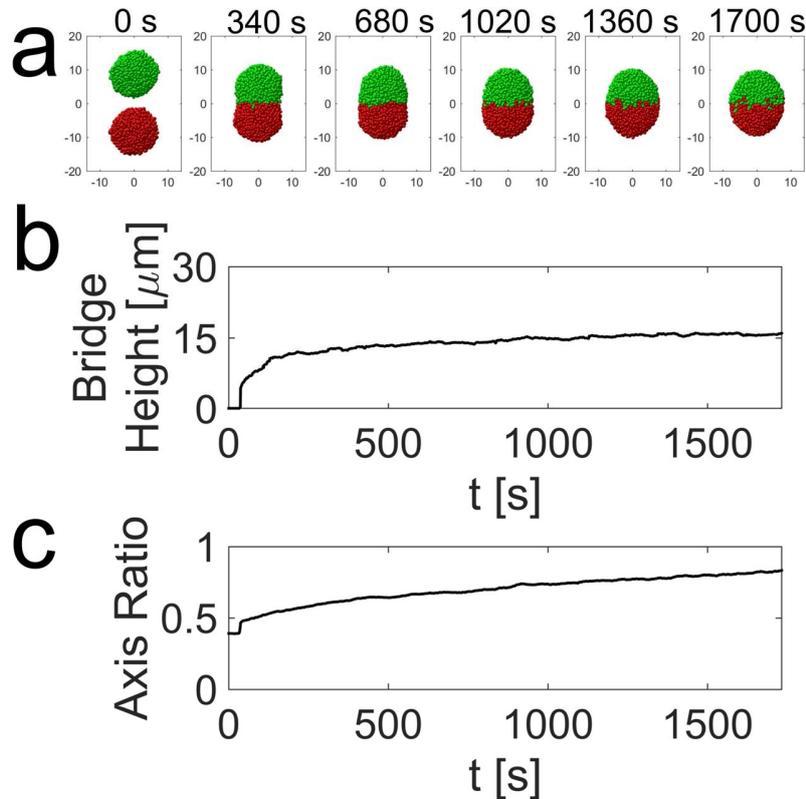

**Figure S4** *In Silico Merger* – **a)** Simulated merger. **b)** Bridge Height of the simulated data **c)** Aspect ratio of the short and long axis for the simulate data.

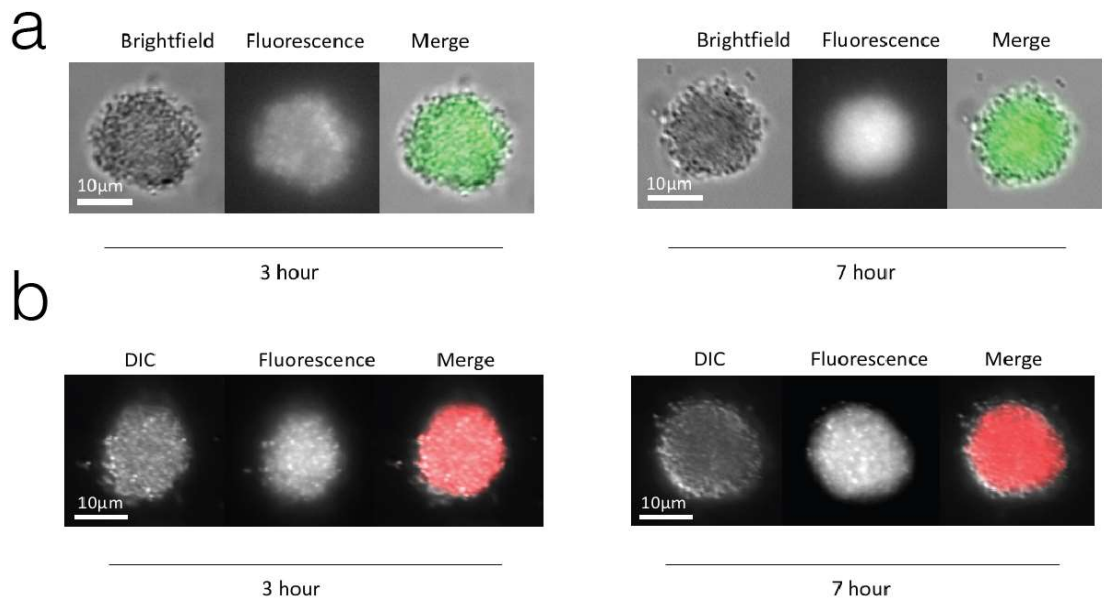

**Figure S5** *Homogeneous metabolic activity in Ng microcolonies*. **a)** Representative fluorescence image of the metabolic dye FDA in a WTNg microcolonies after 3 and 7 hours of formation. Left image is brightfield. Middle image represents the FDA fluorescence channel. The right image is the merge of both (Scale bar = 10 µm). **b)** Representative fluorescence image of the metabolic dye CTC in a WT Ng microcolonies after 3 and 7 hours of formation. Left image is brightfield. Middle image is the CTC fluorescence channel. Right Image is the merge of both (Scale bar = 10 µm)

**Supplementary MovieS1** *Formation of microcolonies by Ng bacteria*. Simultaneous DIC and fluorescence images were taken under a microscope for 3 hours as the bacteria move to form microcolonies mostly by successive merging events of smaller microcolonies (Scale bar = 20 µm).

**Supplementary MovieS2** *Merger of two Ng microcolonies.* The movie represents the merger of two Ng microcolonies where one is represented in green (YFP fluorescence) and the other in red (tdTomato fluorescence).

**Supplementary MovieS3** *Detection and tracking of single cells inside of microcolonies*. While DIC images were recorded with a frequency of 0.1 Hz, fluorescence images were recorded at 1 Hz. The center of mass (COM) of the microcolonies was computed from the DIC images. In order to estimate the position of the COM during the period in which no DIC images were taken, a cubic spline data interpolation applied on the the x- and y-component was used.

# Description of Theoretical Simulation Model

**General description**

The cell body is modeled by two spheres, called cocci, with radius $r$, positions $\vec{r}_1$ and $\vec{r}_2$ and a fixed distance $d < 2r$ between the centers of the spheres to represent the diplococcus shape[1]. Individual pili protrude and retract from a fixed point on the surface of the cell.

**Pili dynamics**

A pilus is characterized by its start and end points $\vec{r}_s$, $\vec{r}_e$ respectively. The contour length of a pilus is given by $l_c = |\vec{r}_e - \vec{r}_s|$. New pili are produced stochastically at a certain rate until a cell has a maximal number of pili $N_{pili}$. The start point of the pilus is randomly distributed on the surface of the cell (see Fig. 1). A new pilus protrudes perpendicular from the cell surface with a constant velocity $v_{pro}$. The pili switch stochastically to the retraction state and retract with the velocity $v_{ret}$. Pili do not re-elongate if retraction started. When the end of a pilus encounters the surface of the substrate, it slides over the surface. For the sake of numerical feasibility, we assume that pili can penetrate through cells. Pili are able to bind to the substrate and form binary connections to pili of other cells. In both situations, a pilus starts to retract immediately after it attached to the surface or another pilus. The binding to the substrate is also stochastic and happens at a certain rate. To model the pili-pili attachment two pili belonging to two neighboring cells are chosen randomly according to the corresponding attachments rates.

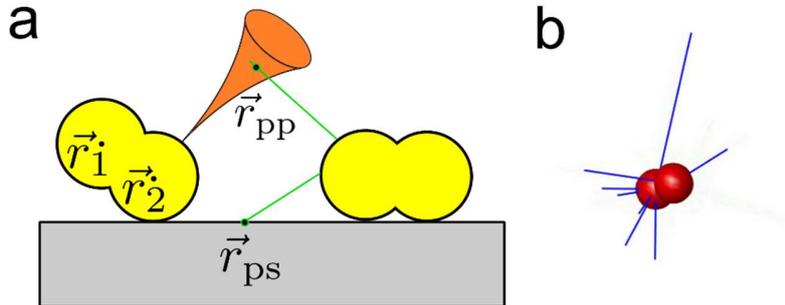

*Figure 1 (a) Pili can bind to the substrate and other pili. (b) Pili of a diplococcus as computed in the simulation.*

To find the coordinates of the intersection point of the two pili one pilus is assumed to sweep a conical region in space, that mimics the thermal fluctuations of an elastic rod with a certain persistence length[2]. The second pilus is modeled as a straight line. The contact point between pili is chosen randomly at the intersection segment of the line and the cone. A sketch of the attachment mechanisms is given in Fig. 1a.

Attachment of the pilus causes its contour length $l_c$ to deviate from its free length $l_f$, which is the length of the freely retracting pilus. While the pilus contour length solely depends on the motion of the cells, its free length is changing in time due to the retraction of the pilus. Pilus is modeled as a Hookean spring[3], which gives magnitude of the pulling force of a pilus attached to the substrate ($F_{ps}$) or another pilus ($F_{pp}$) via the following relation:

$$F_{ps,pp} = \max[0, k_{pull} \cdot (l_c - l_f)]$$

where $k_{pull}$ is the spring constant. Here we assume that the spring constant is high and we can neglect ist dependence on the length of the pili. The retraction velocity of the attached pilus is force dependent:

$$v_{att}(F) = \max\left[0, v_{ret} \cdot \left(1 - \frac{F}{F_{stall}}\right)\right]$$

Here $F$ is the absolute value of the force acting on the pilus, either $\vec{F}_{ps}$ or $\vec{F}_{pp}$. $F_{stall}$ is the stalling force and is a measure for the pulling force of a pilus[4]. The pulling force also affects the detachment probability of the pilus. For the substrate-bound pili and the pili-pili-bonds the detachment rates have the standard force dependent form

$$\lambda(F) = \frac{1}{\tilde{t}} \cdot \exp\left(\frac{F}{\tilde{F}}\right)$$

with $\tilde{t} = t_{d,sub}, t_{d,pp}$ and $\tilde{F} = F_{d,sub}, F_{d,pp}$. Here $F_{d,sub}$ and $F_{d,pp}$ are the detachment forces, $t_{d,sub}$ and $t_{d,pp}$ are the detachment times for the substrate and pili interactions respectively. A detached pilus is able to rebind to the substrate or another pilus with the same rate as a growing pilus.

**Cell Forces and Motility**

The forces in the system cause translation and rotation of the cells. An overlap of the cocci of two different cells (positions $\vec{r}_1$ and $\vec{r}_2$) causes repulsive force

$$F_{cc} = -k_{cc} \cdot (2r - |\vec{r}_{12}|)$$

with $\vec{r}_{12} = \vec{r}_2 - \vec{r}_1$ and the excluded volume spring constant $k_{cc}$. A similar equation describes the contribution of the substrate which is located at $z = 0$, resulting in the force $F_{cs}$.

The total force acting on the center of mass of the cells $\vec{r}_{COM}$ is the sum of the excluded volume forces between the cells $\vec{F}_{cc}$ and the substrate $\vec{F}_{cs}$, the forces of pili attached to the substrate $\vec{F}_{ps}$ and the forces of pili attached to other pili $\vec{F}_{pp}$, so that

$$\vec{F}_{tot} = \sum \vec{F}_{cc} + \sum \vec{F}_{cs} + \sum_{pili} \vec{F}_{ps} + \sum_{pili} \vec{F}_{pp}$$

In the overdamped limit[5], when the viscous drag forces dominate inertia (which holds for the small sizes of bacteria in water) the cell velocity is related to the force by a friction coefficient $\mu_{trans}$:

$$\frac{d}{dr}\vec{r}_{COM} = \frac{d}{dr}\vec{r}_1 = \frac{d}{dr}\vec{r}_2 = \mu_{trans} \cdot \vec{F}_{tot}$$

The mobility results from the Stokes friction of a spherical object in water. The rotation of a cell is described in a similar manner. The total torque is given by

$$\vec{T}_{tot} = \sum (\vec{r}_{cc} - \vec{r}_{COM}) \times \vec{F}_{cc} + \sum (\vec{r}_{cs} - \vec{r}_{COM}) \times \vec{F}_{cs} + \sum_{pili} (\vec{r}_s - \vec{r}_{COM}) \times \vec{F}_{ps}$$

$$+ \sum_{pili} (\vec{r}_s - \vec{r}_{COM}) \times \vec{F}_{pp}$$

Here $\vec{r}_{cc}$ is the contact point of two cocci, $\vec{r}_{cs}$ the contact point between the cell and the substrate, $\vec{r}_s$ is the start point of the pilus and $\vec{r}_{COM}$ is the center of mass of the cell. The angular velocity is proportional to the magnitude of the torque

$$\omega = \mu_{rot} \cdot |\vec{T}_{tot}|$$

Here the rotational mobility $\mu_{rot}$. The angular velocity is used to rotate the cocci positions and the pili start and free pili end points around the center of mass of the cell. The above equations of motion are solved by using the simplest explicit Euler algorithm[6] (we did check, however, that results do not change if we use higher order iteration schemes).

**Parameter Sampling**

Our model has total of 22 parameters, summarized in the Table 1. Importantly, most of parameters are known from literature and only 7 of them (highlighted in the Table 1) were used as free parameters (often with a known admissible range). The parameter set, which showed the best semi-quantitative agreement to the experimental data is provided in the Table I.

| Parameter | Symbol | Value |
|---|---|---|
| Time step [s] | $\Delta t$ | $5 \cdot 10^{-6}$ |
| Cocci radius [µm] | $r$ | $0.5^1$ |
| Cocci distance [µm] | $d$ | $0.6^1$ |
| Cell-Cell excl. vol. constant [pN/µm] | $k_{cc}$ | $10^4$ |
| Cell-Sub excl. vol. constant [pN/µm] | $k_{cs}$ | $2 \cdot 10^4$ |
| Transl. friction coeff. [µm/(pN · s)] | $\mu_{trans}$ | 1 |
| Rotat. friction coeff. [1/(pN · s)] | $\mu_{rot}$ | 2 |
| Characteristic pili length [µm] | $l_{ch}$ | $1.5^1$ |
| Pili persistence length [µm] | $l_p$ | $5.0^7$ |
| Pili production rate [1/s] | $\lambda_0$ | 15 |
| *Maximal pili number | $N_{pili}$ | $15^{1,8}$ |
| Pili protrusion velocity [µm/s] | $v_{pro}$ | $2^1$ |
| Pili retraction velocity [µm/s] | $v_{ret}$ | $2^1$ |
| Pili retraction rate [1/s] | $\lambda_{ret}$ | $1.33^1$ |
| *Pili substrate attachment rate [1/s] | $\lambda_{sub}$ | 0.5 |
| *Pili Pili attachment rate [1/s] | $\lambda_{pil}$ | 0.5 |
| Pili spring constant [pN/µm] | $k_{pull}$ | $2000^3$ |
| Stalling force [pN] | $F_{stall}$ | $180^8$ |
| *Substrate detachment time [s] | $t_{d,sub}$ | 10 |
| *Substrate detachment force [pN] | $F_{d,sub}$ | 180 |
| *Pili-Pili detachment time [s] | $t_{d,pp}$ | 50 |
| *Pili-Pili detachment force [pN] | $F_{d,pp}$ | 360 |

TABLE I. Parameter set used if not stated otherwise. The stars mark the parameters for which we performed search for optimal values. All other parameters were taken from previous studies.

## Single Colonies and Merger

To find the best matching parameter set we simulated 30 minutes worth of the merging and single colonies for each parameter set, given in Table II.

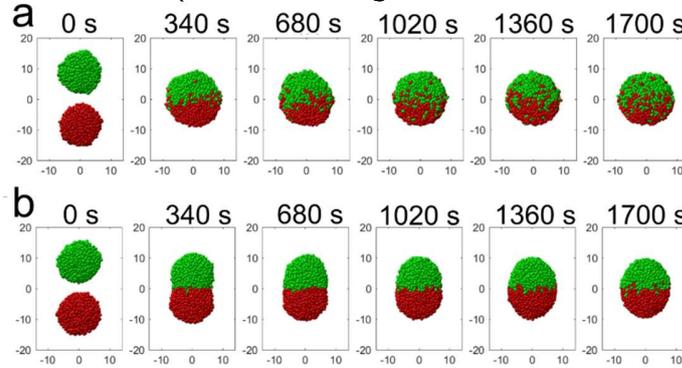

Figure 2 Example of a merger for two different parameter sets: (A) $N_{pili} = 15$, $\lambda_{pil} = 0.5$, $t_{d,pp} = 20$ s, $F_{d,pp} = 180$ pN  (B) $N_{pili} = 15$, $\lambda_{pil} = 2$, $t_{d,pil} = 50$ s, $F_{d,pil} = 360$ s

| Cells – Single Colony | $N$ | 1700 |
|---|---|---|
| Cells - Merger | $N$ | 1000+1000 |
| *Maximal pili number | $N_{pili}$ | 10, 15 |
| *Pili Pili attachment rate [1/s] | $\lambda_{pil}$ | 0.25, 0.5, 2 |
| *Pili-Pili detachment time [s] | $t_{d,pp}$ | 5, 20, 30, 40, 50.60, 70 |
| *Pili-Pili detachment force [pN] | $F_{d,pp}$ | 120, 180, 240, 300, 360 |

TABLE II. Parameter range for the sampling of single colony and merger simulations. While low detachment times, detachment forces, pili numbers and attachment rates enhanced the motility of the cells inside of the colonies and accelerated the merging, higher values reduced the motility and slowed down the merging (see Fig. 2).

## Assembly

By estimating four of seven free parameters from the merger and single colony simulations we performed simulations of the assembly of cells on a substrate for the remaining three free parameters (see table III). For every set of parameters one simulation was performed, as shown in Fig. 3, reproducing a total time of 1.5 hours. For a wide range of parameters we observe the demixing of ∆pilT mutants and WT cells.

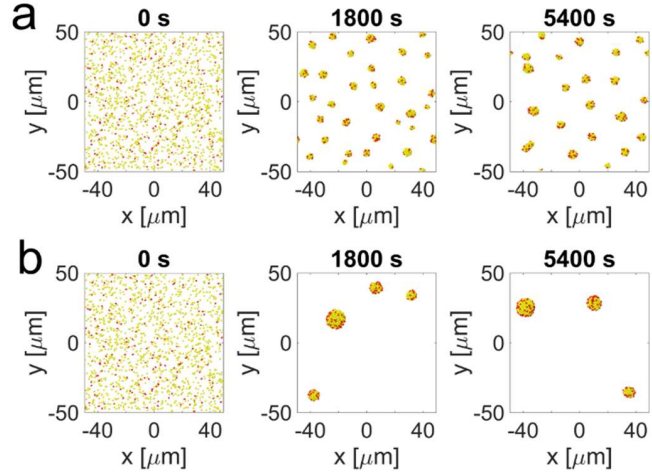

*Figure 3 Example of assembly for two different parameter sets: (A) $N_{pili} = 15$, $\gamma = 0.1$, $\lambda_{sub} = 0.5$, $t_{d,sub} = 60$ s, $F_{d,sub} = 300$ s (B) $N_{pili} = 15$, $\gamma = 0.1$, $\lambda_{sub} = 0.5$, $t_{d,sub} = 10$ s, $F_{d,sub} = 180$ s*

| Cells Number | $N$ | 1500 |
|---|---|---|
| Box Size [μm×μm] | | 98.56×98.56 |
| Percentage ΔpilT | $\gamma$ | 0, 0.1, 05 |
| *Pili substrate attachment rate [1/s] | $\lambda_{sub}$ | 0.25, 0.5, 2 |
| *Substrate detachment time [s] | $t_{d,sub}$ | 0.1, 10, 30, 60 |
| *Substrate detachment force [pN] | $F_{d,sub}$ | 2, 180, 300, 400 |

*TABLE III. Parameter range for the sampling of single colony and merger simulations.*

**Comment on the agreement between experiments and model:**
Currently our model provides a semi-quantitative agreement with the experimental data. Most likely the discrepancy is due to the simplification of the model allowing for one pilus to have at maximum one contact point with any other pili of other cells. While dictated by computer feasibility it leads to more "dynamic" microcolonies as compared to experiments. Multiple pili contact points in the real setting lead to the formation of active pili network with higher gradient in motility and less dynamic core of the cells. It is the interesting direction of further research to understand the biophysical properties of such a network.